\documentclass[preprint2]{aastex}

\newcommand{\velunit}{\,\rm km\,s^{-1}}
\newcommand{\HeI}{\ion{He}{1}}
\newcommand{\OII}{[\ion{O}{2}]}
\newcommand{\OIII}{[\ion{O}{3}]}

\newcommand{\ob}{12+\rm log(O/H)}

\shorttitle{Spectroscopy of Young Star Clusters in NGC\,4676}
\shortauthors{Chien et al.} 

\begin{document}

\title{Multi-object Spectroscopy of Young Star Clusters in NGC\,4676\footnote{The data presented herein were obtained at the W.M. Keck Observatory, which is operated as a scientific partnership among the California Institute of Technology, the University of California and the National Aeronautics and Space Administration. The Observatory was made possible by the generous financial support of the W.M. Keck Foundation.}}
\author{Li-Hsin Chien, Joshua E. Barnes, Lisa J. Kewley, Kenneth C. Chambers}
\affil{Institute for Astronomy, University of Hawaii, 2680 Woodlawn Drive, Honolulu, HI 96822}
\email{chien@ifa.hawaii.edu}
\email{barnes@ifa.hawaii.edu}

\begin{abstract}
Galaxy interactions are known to trigger starbursts.  The young star clusters formed in mergers may be young globular clusters.  The ages of these young star clusters yield the timing of interaction-triggered star formation and provide an important way to reconstruct the history of merging galaxies.  Here we present the first results from our investigation into age and metallicity of twelve young clusters in the merging galaxy pair NGC\,4676, using spectra from the multi-object Low Resolution Imaging Spectrometer (LRIS) on Keck.  For ten clusters, comparison of the Balmer emission lines with model equivalent widths (EWs) yields ages less than 10\,Myr.  Two spectra display Balmer absorption lines typical of star clusters dominated by A-type stars, with estimated ages of about 170\,Myr.  These ages are comparable to the dynamical age of the tidal tails and are consistent with star formation triggered during the first passage of the pair.  The locations of these two clusters in the tidal tails are generally consistent with predictions of shock-induced star formation models.  One of these older objects appears unresolved on the image and is luminous enough to qualify as a young globular cluster.  Using EWs of the diagnostic lines \OII\ and \OIII, we obtain oxygen abundances in the range $7.3<\ob<9.0$.  These values show a nearly flat distribution along the northern tail, suggesting efficient gas mixing in the tail.

\end{abstract}


\section{INTRODUCTION}
Gas-rich merging galaxies, e.g.\,NGC\,4038/39 \citep[][etc]{whitmore99} and NGC\,3256 \citep{zepf,trancho}, are now known to host numerous young massive star clusters that formed during the merger.  These young star clusters formed during an interaction-induced starburst can provide a powerful diagnostic for tracing star formation history \citep[e.g.][]{schweizer}.    With the Mice as a starting point, our goal is to use multi-object spectroscopy to map the ages of young clusters throughout the bodies and tails of interacting galaxies, and thereby trace the history of star formation during the encounter.  

The two galaxies making up NGC\,4676 were dubbed ``the Playing Mice" by \citet{voron}.  The pair is located at $\alpha_{\rm J2000.0} = 12^{h}43^{m}44^{s}$, $\delta_{\rm J2000.0} = +31\arcdeg00\arcmin18\arcsec$ \citep{rines} and has a recession velocity $cz_{\sun} = 6610 \velunit$.  Adopting a Hubble constant of $H_{0} = 72 \velunit {\,\rm Mpc^{-1}}$ places the Mice at a distance of 88.8\,Mpc, corresponding to a scale of $1\arcsec = 0.43 {\,\rm kpc}$.  A dynamical simulation suggested that the first pericentric passage of the pair occurred about 170\,Myr ago, and that the extended star formation observed in the system is better described by shock-triggered star formation \citep{barnes}.  To test the validity of this model, we obtained spectra of twelve young star clusters and associations in the Mice.  Here we present the spectra and first results on the ages and metallicities of these clusters.

\section{OBSERVATION AND REDUCTION}
Observations were performed on 2005 May 4, with LRIS \citep{oke} on Keck I.  The target star clusters were chosen from the brighter candidates in \citet{degrijs}; we attempted to sample clusters from every part of the color-color diagram (see their Figure 4).  The astrometric solution in \cite{degrijs} was based on the Hubble Space Telescope (HST) images, producing an intrinsic scale error of about 1\% in the tabulated coordinates of these clusters.  Corrections to the coordinates were made using the USNO-A2.0 catalogue \citep{monet} as an absolute astrometric standard.  Table 1 shows the corrected coordinates and measured properties of each cluster.  A {\it g}-band image of the Mice taken by the Advanced Camera for Surveys \citep[ACS;][]{ford} and the position of slits and clusters in the Mice are shown in Figure 1.  Each slit width is 1$\arcsec$, with lengths ranging from $5\arcsec {\ \rm to}\ 16\arcsec$.  Two slits captured spectra of additional off-target stellar populations and star clusters.  The blue spectra were obtained with a dispersion of 1.09\,\AA\ per pixel, a spectral resolution of 5.95\,\AA, and an approximate wavelength coverage of 3200\,\AA\ to 5600\,\AA.  Nine 30 minute exposures were taken, together with standard star observations and short calibration exposures.  Five of the exposures were partly obstructed by passing cirrus, thus only four frames were used to obtain the spectra analysis.

Spectral reduction was performed using standard IRAF\footnote{IRAF is distributed by the National Optical Astronomy Observatory (NOAO), which is operated by the Association of Universities for Research in Astronomy (AURA), Inc. under a cooperative agreement with the National Science Foundation (NSF).} tasks.  The blue data frames were bias subtracted, registered, combined, cosmic ray cleaned, distortion corrected and wavelength calibrated.  Relative flux calibration was performed using three standard stars; absolute calibration was not attempted since conditions were non-photometric and obtaining EWs only requires relative calibration.  

Sky and galaxy background subtractions were performed seperately for each of the clusters.  Two slits (not shown in Figure 1) in our slitmask were configured to obtain the sky background; averaging these and boxcar smoothing them in the spatial direction gave us a single sky background image.  The sky background image was then spatially scaled and subtracted from each of the cluster images.  Next, the one-dimentional spectra of the clusters were extracted with an aperture of about 1$\arcsec$.  Two adjacent apertures of the same width were simultaneously extracted and combined to yield the background galaxy spectrum.  Both the cluster spectrum and the background galaxy spectrum were smoothed by 9 pixels ($= 9.8 {\,\rm \AA}$).  The continuum and the emission spectra of the clusters were found to be clearly defined, and the galaxy background had no steep gradient for each of the clusters, thus allowing a simple linear subtraction.  Visual inspection suggests that this subtraction effectively removed the galaxy background from the cluster spectra.  Figure 1 shows the spectra of the clusters, plotted as relative flux versus observed wavelength.  Note that cluster candidate S3, or No.\,32 in \citet{degrijs}, is a foreground star.

\section{SPECTROSCOPIC RESULTS}
\subsection{Ages of Star Clusters in NGC\,4676}
The spectra of most clusters show a nearly flat continuum with a combination of strong nebular emission and underlying stellar absorption features, suggesting populations of very young massive stars surrounded by ionized gas.  Two of the spectra, cluster S1 and S9a, show Balmer absorption lines along with relatively weak \ion{Ca}{2} K lines, indicating that these clusters are dominated by A-type stars \citep{schweizer98}.  The Balmer jump and the prominent optical continuum of these two clusters are signs of faded young massive stars and older ages.     

The Balmer lines (H$\beta$, H$\gamma$, H$\delta$, H8 and H9) were measured using the SPLOT and NGAUSSFIT tasks in IRAF.  In order to separate the emission and the absorption features of some Balmer lines, both features were simultaneously fit with Gaussian profiles.  The equivalent widths (EWs) were then calculated and redshift corrected, i.e. EW$_{rest}$ = EW$_{observed}$/(1+z).  Age estimates for the clusters were derived by comparing the measured EW of each line with the synthesized EW from \citet[][hearafter GLH99]{gon}, assuming a Salpeter IMF and solar metallicity.  Note that the nebular emission of Balmer series were used to derive ages less than 20 Myr (c.f.\,Table 7 in GLH99), and for cluster S1 and S9a the absorption features were used (c.f.\,Table 1 in GLH99).  Averaging estimates for different lines yielded the age and the statistical error for each cluster shown in Table 1.  

Although the signal to noise ratio was not high enough to detect most \HeI\,lines, \HeI\,$\lambda4471$ was observable in five of the spectra.  The spectra of cluster S2, S5 and S10 showed emission features, whereas cluster S7 and S8 showed only absorption features.  The EWs of the \HeI\,$\lambda4471$ line were measured using SPLOT, redshift corrected and compared with the synesized EWs using the same method described above (c.f.\,Table 3 and Table 7 in GLH99).  Age estimates based on this line and the calculated uncertainties from propogation of errors are also shown in Table 1.  The age estimates from Balmer lines and \HeI\,$\lambda4471$ line are in a good agreement with each other.

\subsection{Metallicity Results}
We adopt the chemical analysis method of \citet[][hereafter KK04]{kobul} to determine metallicity.  The EW ratio of the collisionally excited \OII$\lambda3727$, \OIII$\lambda\lambda4959,\,5007$ emission lines relative to the H$\beta$ recombination line, known as R$_{23}$, allows an estimate of gas-phase oxygen abundances, $\ob$, of the young clusters.  Instead of traditional flux ratio, this method using EW ratios has the advantage of being reddening independent to first order.  The EWs of the \OII\ and \OIII\ lines were measured using the SPLOT task in IRAF and were redshift corrected.  The EWs of H$\beta$ used for the following metallicity calculation were the same as those used for the age estimates, where the emission features were corrected from the absorption features.   

The oxygen abundance, however, is known to be degenerate with the R$_{23}$ diagnostic line ratio, giving both a high ``upper branch" and a smaller ``lower branch" metallicity estimate (e.g.\,Figure 7 in KK04; the separation between the two branches occurs at 12\,+\,log[O/H]\,$\sim8.4$).  In this work we calculated the oxygen abundances for both branches using the analytical expressions of McGaugh(1991) in KK04\footnote{Equation 15; also see \citet{kobul99} Equation 8 for expressions of oxygen abundances on the lower branch.}.  To break the degeneracy, we used the EW of the \OII\,$\lambda3727$ line as the discriminator.  L. J. Kewley \& H. A. Kobulnicky (2007, in preparation) suggest that the oxygen abundance would have values represented by the upper branch if the EW(\OII)\,$\la$\,50\,\AA\ or by the lower branch if the EW(\OII)\,$>$\,200\,\AA.  Note that for cluster S10, the EW(\OII) lies in between these limits so that its oxygen abundance is still uncertain.  In Table 1 we list the oxygen abundances of each cluster under the ``lower" and ``upper" columns, followed by the estimated uncertainties due to the H$\beta$ absorption features.  These uncertainty estimates do not include the error introduced by the model uncertainties, typically $\sim$ 0.1 dex, in the theoretical calibrations.  Our best guess for the metallicity, depending upon which branch it lies on, is listed in the ``adopted" column of Table 1.  

For most of the clusters, their adopted oxygen abundances are greater than 8.91.  This value corresponds to a metallicity of $1.6\,Z_{\odot}$\footnote{The metal mass fraction, $Z$, is related to the oxygen a bundance by $Z\simeq29\times10^{[12+log(O/H)]-12}$, for the standard solar abundance distribution with the solar oxygen abundance of \citet{allende}, which yields a solar metallicity of $Z_{\odot}\simeq0.015$ and 12\,+\,log(O/H)$_{\odot}=8.72$.}.  Cluster S2 has an adopted oxygen abundance of 7.34, which corresponds to a metallicity of $0.04\,Z_{\odot}$.  These are distinctly non-solar metallicities, however our age estimates were based upon a model assuming a solar value.  Therefore corrections to the derived ages needed to be considered.  The effect that these metallicities have on our age estimates is manifested in the evolution of the H$\beta$ EW \citep{leitherer}.  For ages less than 500\,Myr, the main-sequence stars that dominate the luminosity have lower effective temperature at high metallicity; consequently the nebula produces weaker H$\beta$ emission lines.  Comparing to the results of \citet{leitherer} our age estimates for clusters with higher metallicity were overstimated by approximately 1 Myr.  In contrast, the low metallicity for cluster S2 produces an underestimate for the age by about 1 Myr.

\section{DISCUSSION}
We obtain reliable age determinations from our spectra of the star clusters in the Mice, and as a collateral result calculate their metallicities for the first time.  \citet{kewley} suggests that low nuclear metallicities in close pairs, atypical of the metallicity gradients of late-type spiral galaxies, may be evidence for tidally induced gas flows.  The galaxy interactions cause gas flows toward the central regions, delivering less enriched gas from the outskirts of the galaxy.  Our metallicities show a nearly flat distribution along the northern tail of the Mice, consistent with \citet{kewley} scenario.  This flat metallicity distribution may suggest gas mixing within the tidal tail.

\citet{barnes} has performed numerical simulations using two star formation rules which yield markedly different predictions for star formation in the Mice.  A shock-induced prescription predicts a widespread burst of star formation triggered by large-scale shocks during the first passage $175 \pm 25$ Myr ago and a second burst of star formation, mostly concentrated within the central regions of the galaxies, developing more recently.  In contrast, a simulation using a density-dependent prescription predicts no global burst of star formation associated with the passage, but rather a gradual increase within the galactic centers.  From our results, most young clusters formed fairly recently ($<$\,10\,Myr), however two were found to have ages of $\sim$\,170\,Myr, which suggests that they most likely formed during the first passage of the Mice.  These two older objects are located in the tidal tails of the pair, which is consistent with the spatial distribution of star formation predicted by shock-induced models.

Young star clusters formed in mergers may evolve into globular clusters, illuminating the origin of the high specific frequency of globular clusters in elliptical galaxies \citep{schweizer86, ashman}.  Cluster S1 appears unresolved on the image, and with an estimated initial mass of 10$^{6}$ M$_{\sun}$ \citep[assuming a Salpeter IMF with low mass cut-off of 0.15 M$_{\sun}$;][]{leitherer}, the cluster is massive and luminous enough to qualify as a young globular cluster.  

\section{CONCLUSION}
Here we present the first results on the ages and metallicities of young clusters in the Mice.  These ages are obtained by comparing the EWs of the Balmer lines to model values, and the oxygen abundances are calculated using standard line diagnostics.  Most of the cluster spectra show strong Balmer emission lines, implying estimated ages from 2.5 to 7.1\,Myr.  Two of the spectra lack emission features, yielding estimated ages of 163 and 172 Myr.  These ages are comparable to the dynamical age of the tidal tails and are consistent with shock-induced star formation triggered during the first passage of the Mice.  The calculated oxygen abundances for the clusters along the northern tidal tail may indicate gas mixing within the tail.  Further detailed comparison with the predicted age distributions of the star clusters from simulations will be presented in a future paper. 

\acknowledgments
We thank Gregory Wirth for expert assistance at the telescope, and the anonymous referee for useful comments that helped improve the manuscript.  The authors recognize and acknowledge the very significant cultural role and reverence that the summit of Mauna Kea has always had within the indigenous Hawaiian community, and are grateful to have had the opportunity to conduct observations from this mountain.  L.-H. C. gratefully acknowledge the support from the STScI through grant HST-GO-09822.07-A, P.I. D.B. Sanders.

\clearpage

\begin{deluxetable}{lcrrcccccc}
\tabletypesize{\scriptsize}
\tablecaption{Property, Estimated Age, and Metallicity of Star Clusters in NGC\,4676}
\tablewidth{0pt}
\tablehead{
\colhead{}&\colhead{Other}&\colhead{RA (J2000)}&\colhead{DEC (J2000)}&\colhead{F475W\tablenotemark{a}}&\colhead{Age\tablenotemark{b}}&\colhead{Age\tablenotemark{c}}&\multicolumn{3}{c}{log(O/H) + 12\tablenotemark{d}}\\
\colhead{No.}&\colhead{ID\tablenotemark{a}}&\colhead{(hh mm ss.ss)}&\colhead{(dd mm ss.ss)}&\colhead{(HST mag)}&\colhead{(Myr)}&\colhead{(Myr)}&\colhead{Lower}&\colhead{Upper}&\colhead{Adopted}}
\startdata
S1\tablenotemark{e}&40&12 46 12.15&30 43 06.86&22.51&172$\pm$33&\nodata&\nodata&\nodata&\nodata\\
S2&35&12.33&25.76&23.26&2.5$\pm$0.54&2.6$\pm$0.36&7.34$\pm$0.03&8.98$\pm$0.02&7.34$\pm$0.03\\
S3\tablenotemark{f}&32&09.73&21.00&23.55&\nodata&\nodata&\nodata&\nodata&\nodata\\
S4&30&11.53&36.00&22.52&5.5$\pm$0.06&\nodata&7.47$\pm$0.02&8.95$\pm$0.01&8.95$\pm$0.01\\
S4a&31&11.57&34.46&21.80&6.0$\pm$0.20&\nodata&7.56$\pm$0.03&8.91$\pm$0.01&8.91$\pm$0.01\\
S5&26&10.18&48.62&19.68&4.1$\pm$0.08&4.8$\pm$1.94&7.33$\pm$0.01&8.96$\pm$0.01&8.96$\pm$0.01\\
S6&24&10.16&55.37&20.15&5.9$\pm$0.09&\nodata&7.76$\pm$0.17&8.83$\pm$0.11&8.83$\pm$0.11\\
S7&8&10.51&30 44 34.75&20.34&6.6$\pm$0.42&6.6$\pm$3.15&7.53$\pm$0.03&8.92$\pm$0.01&8.92$\pm$0.01\\
S8&6&10.51&53.30& 20.66&7.1$\pm$0.10&7.7$\pm$2.39&7.43$\pm$0.16&8.97$\pm0.07$&8.97$\pm$0.07\\
S9&5&10.57&30 45 16.42&21.68&5.9$\pm$0.04&\nodata&7.58$\pm$0.11&8.92$\pm$0.06&8.92$\pm$0.06\\
S9a&\nodata&10.62&14.09&\nodata&163$\pm$35&\nodata&\nodata&\nodata&\nodata\\
S9b\tablenotemark{e}&\nodata&10.65&12.21&\nodata&5.6$\pm$0.11&\nodata&7.62$\pm$0.04&8.96$\pm$0.02&8.96$\pm$0.02\\
S10&2&10.56&31.62&22.97&2.7$\pm$0.22&2.9$\pm$0.27&7.52$\pm$0.10&8.86$\pm$0.04&\nodata
\enddata
\tablenotetext{a}{Data from \citet{degrijs}.}
\tablenotetext{b}{Derived by comparing EWs of Balmer lines with GLH99, assuming solar metallicity.}
\tablenotetext{c}{Derived by comparing EWs of He I\,$\lambda$4471 line with GLH99, assuming solar metallicity.}
\tablenotetext{d}{Calculated from KK04 and Kobulnicky et al.\,(1999).}
\tablenotetext{e}{No Oxygen nebular lines were observed.}
\tablenotetext{f}{A foreground star.}
\end{deluxetable} 

\clearpage

\onecolumn
\begin{figure}
\plotone{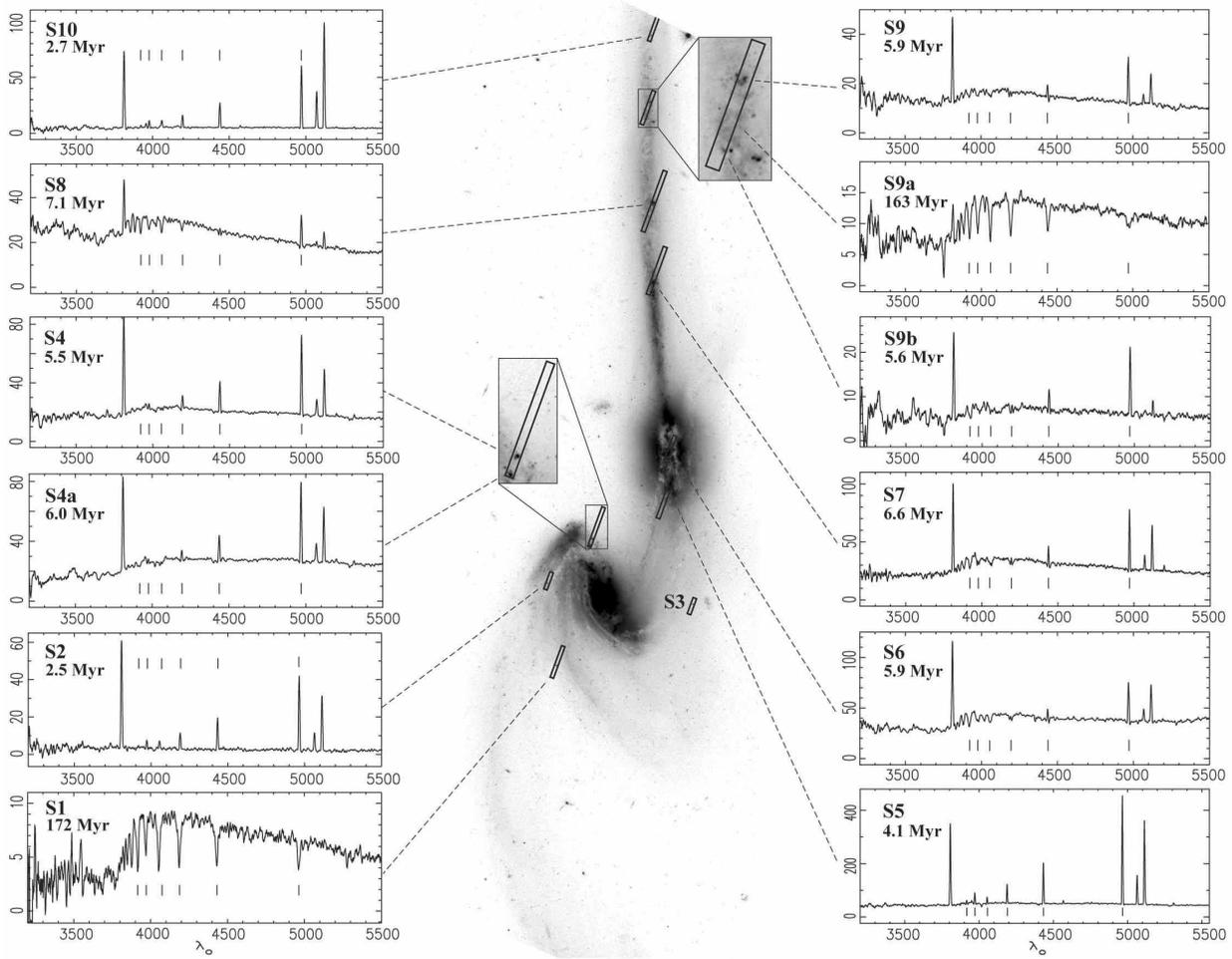}
\caption{ACS {\it g}-band image of the Mice, with blow-ups of star cluster S9, S9a and S9b on the right and S4 and S4a on the left.  Boxes represent the slits used in the spectroscopy.  Age of each star cluster is shown with its spectrum, plotted as relative flux vs. observed wavelength (\AA).  Markers in the spectra are the Balmer series.  Note that S3 is a foreground star.}
\end{figure}

\end{document}